\documentclass{article}
 
\usepackage{fullpage, amsmath, amssymb, amsthm}
\usepackage{xspace}
\newcommand{\F}{\mathbb{F}}
\newcommand{\Fam}{\mathcal{M}}
\newcommand{\D}{\mathcal{D}}
\newcommand{\E}{\mathbb{E}}
\newcommand{\Err}{\mathcal{E}}
\newcommand{\Patrascu}{P{\v a}tra{\c s}cu\xspace}

\newtheorem{theorem}{Theorem}
\newtheorem{lemma}{Lemma}
\newtheorem{definition}{Definition}

\DeclareMathOperator{\vspan}{span}
\newcommand{\RankSum}{\mathcal{RS}}

\title{New Unconditional Hardness Results for Dynamic and Online Problems}

\author{Raphael Clifford\thanks{Bristol University. \texttt{Raphael.Clifford@bristol.ac.uk}} \and Allan Gr\o nlund\thanks{Aarhus University. \texttt{jallan@cs.au.dk}. Supported by Center for Massive Data Algorithmics, a Center of the Danish National Research Foundation, grant DNRF84.}
  \and Kasper Green Larsen\thanks{Aarhus University. \texttt{larsen@cs.au.dk}. Supported by Center for Massive Data Algorithmics, a Center of the Danish National Research Foundation, grant DNRF84.}}

\date{}

\begin{document}
\maketitle
\begin{abstract}
 There has been a resurgence of interest in lower bounds whose truth rests on the conjectured hardness of well known computational problems. These conditional lower bounds have become important and popular due to the painfully slow progress on proving strong unconditional lower bounds. Nevertheless, the long term goal is to replace these conditional bounds with unconditional ones. In this paper we make progress in this direction by studying the cell probe complexity of two conjectured to be hard problems of particular importance: matrix-vector multiplication and a version of dynamic set disjointness known as \Patrascu's Multiphase Problem.  We give improved unconditional lower bounds for these problems as well as introducing new proof techniques of independent interest. These include a technique capable of proving strong threshold lower bounds of the following form: If we insist on having a very fast query time, then the update time has to be slow enough to compute a lookup table with the answer to every possible query. This is the first time a lower bound of this type has been proven.
\end{abstract}

\thispagestyle{empty}
\setcounter{page}{0}
\newpage
\pagestyle{plain}
\section{Introduction}

Proving lower bounds for basic computational problems is one of the most challenging tasks within computer science. Where optimal bounds can often be found for space requirements, we are still a long way from being able to establish similar results for time complexity for all but a relatively small subset of the problems we wish to study.  Due to the difficulty in obtaining these lower bounds, in recent years there has been a resurgence in interest in finding bounds which hold conditioned on the conjectured hardness of a small number of widely studied problems.  Perhaps the most prominent examples are 3SUM-hardness (see e.g.\@~\cite{GO:1995}), reductions from the Strong Exponential Time Hypothesis (SETH)~\cite{PW:sat:2010, RW:sparse:2013, AW:2014,AWW:2014, BI:SETH:2015, Bringmann:2014,BK:TimeWarping:2015} and specifically for dynamic problems, reductions from a version of dynamic set disjointness known as \Patrascu's Multiphase Problem~\cite{Patrascu:2010:polylower} and most recently online Boolean matrix-vector multiplication~\cite{HKNS:2015}. Of course the holy grail remains to prove strong unconditional lower bounds for these problems. Unfortunately the state-of-the-art techniques for proving lower bounds for data structure problems such as Boolean matrix-vector multiplication can only prove time lower bounds of $\Omega(\lg m)$, where $m$ is the number of queries to the problem. For the online Boolean matrix-vector multiplication problem there are $2^n$ queries, which means we cannot hope to prove bounds beyond $\Omega(n)$ without ground breaking new insight. This is quite disappointing given that the conjectured complexity of the problem is $n^{2-o(1)}$.

In this paper we add to the understanding of the true complexity of dynamic and online problems by giving new unconditional lower bounds for \Patrascu's Multiphase Problem as well as online and dynamic matrix-vector multiplication over finite fields. Our focus is to prove unconditional polynomial lower bounds for restricted ranges of trade-offs between update time, query time and space. 

For \Patrascu's Multiphase Problem, we prove a new type of threshold lower bound saying that if we insist on having a very fast query time, then the update time essentially has to be high enough to compute a lookup table of the answer to every possible query. This is the first threshold lower bound of this form.

For matrix-vector multiplication, the lower bounds we prove demonstrate that if a data structure doesn't explicitly try to exploit that it is dealing with a small finite field, then it is doomed to spend $n^{2-o(1)}$ time per operation. Furthermore, our lower bounds are as strong as current techniques allow. Matrix-vector multiplication is a basic computational primitive in applied mathematics and so our new bounds for this problem are also of separate and independent interest. 

The lower bounds we prove are all in the cell probe model of computation. We present this model in the following.

\subsection*{Cell probe model}

A data structure in the cell probe model consists of a set of memory cells, each storing $w$ bits. Each cell of the data structure is identified by an integer address, which is assumed to fit in $w$ bits, that is each address is amongst $[2^w] = \{0,\dots, 2^w-1\}$. So that a cell has enough bits to address any update operation performed on it, we will assume $w \in \Omega(\lg n)$ when analysing a data structure's performance on a sequence of $n$ updates.

During an update operation, the data structure reads and updates a number of the stored cells to reflect any changes. The cell read or written to in each step of an update operation may depend arbitrarily on both the update and the contents of all cells previously probed during the update. The update time of a data structure is defined as the number of cells probed, that is read or written to, when processing an update. 

In order to answer a query, a data structure probes a number of cells from the data structure. From the contents of the probed cells, the data structure must return an answer to the query. As with update operations, the cell probed at each step, and the answer returned, may be an arbitrary function of the query and the previously probed cells. We define the query time of a data structure as the number of cells probed when answering a query.  

The cell probe model was introduced originally by Minsky and Papert~\cite{MP:1969}  in a different context and then subsequently by Fredman~\cite{Fredman:1978}  and Yao~\cite{Yao1981:Tables}. The generality of the cell probe  model makes it particularly attractive for establishing lower bounds for dynamic  data structure problems.   The cell probe model, for example, subsumes the popular word-RAM model.

\subsection*{Previous cell probe lower bounds}

The main approaches for proving dynamic data structure lower bounds in the cell probe  model have historically been based on the chronogram
technique of Fredman and Saks~\cite{FS1989:chronogram}, which until approximately a decade ago was able to prove $\Omega(\lg{n}/\lg{\lg{n}})$ lower
bounds at best. This technique was based on partitioning a sequence of updates into \emph{epochs} of geometrically decreasing size and then arguing that, amongst the cells updated during each epoch, any correct data structure has to probe $\Omega(1)$ of them. In 2004 a breakthrough led by \Patrascu and Demaine developed the \emph{information transfer technique} which gave the first $\Omega(\log{n})$ lower bound per operation for several data structure problems~\cite{PD2006:Low-Bounds}. Later on it was also shown that an $\Omega(\lg{n})$ time lower bound can be derived using the same approach for the related questions of streaming and online computation, including multiplication and various string matching problems~\cite{CJ:2011,CJS:2013, CJS-soda:2015}.  The key difference between the streaming and online problems and a standard dynamic data structure setting is that although there are still many different possible updates at each step, there is only one query which is simply to output the latest result.

In 2012 there was another breakthrough for dynamic data structure lower bounds.  The new idea was to combine the cell sampling approach of Panigrahy et al.\@~\cite{PTW:2010} with the chronogram technique of Fredman and Saks. In essence, this approach allows one to argue that when answering a query, one has to probe $\Omega(\lg{m}/\lg(wt_u))$ cells from each epoch instead of $\Omega(1)$. With around $\lg n/\lg(w t_u)$ epochs, this gives lower bounds of roughly $\Omega(\lg n \lg m / (\lg(w t_u))^2)$. Here $m$ is the number of queries that can be asked in the data structure problem. This resulted in an $t_q = \Omega((\lg{n}/\lg(wt_u))^2)$ lower bound for dynamic weighted range counting and $t_q =  \Omega(\lg{|\F|}\lg{n}/\lg(wt_u/\lg{|\F|})\lg(wt_u))$ for dynamic polynomial evaluation when computing over a field $\F$ of size at least $\Omega(n^2)$~\cite{Larsen:2012,Larsen:2012:focs}. This latter bound was, until this current work, the only such bound that holds for randomised data structures which can err with constant probability. Perhaps due to the technical difficulties involved, no further lower bounds of this form have been shown to date.

Attacking the problem of finding lower bounds from a different angle, \Patrascu and Thorup showed a sharp query/update time trade-off for dynamic connectivity in undirected graphs. They showed that any data structure that supports edge insertions in $o(\lg n)$ probes, must have worst case connectivity time $n^{1-o(1)}$ in the cell probe model assuming cells of $\Theta(\lg{n})$ bits~\cite{PT:stoc:2011}. In other words, really fast updates imply nearly naive running time for queries. 

Towards the aim of giving yet higher lower bounds, in~\cite{Patrascu:2010:polylower} \Patrascu introduced a dynamic version of set disjointness which he termed the Multiphase Problem.  He showed reductions for this problem, first from 3SUM and then to dynamic reachability, dynamic shortest path as well as subgraph connectivity and other problems of general interest.  Assuming  that there is no truly sub-quadratic time solution for 3SUM, he was then able to give the first known polynomial time lower bounds for many dynamic data structure problems. 

Online matrix-vector multiplication~\cite{HKNS:2015} can also be viewed as a static problem in classic data structure terminology, that is we receive some data to preprocess (a matrix) and then we answer queries (vectors). Thus we find it relevant to also list previous techniques and barriers for proving static cell probe lower bounds. 

One of the early techniques for proving static lower bounds was based on a reduction from asymmetric communication complexity by Miltersen et al.~\cite{MNSW:1998}. This technique led to lower bounds of the form $\Omega(\lg m/ \lg S)$ where $m$ is the number of queries in the data structure problem and $S$ is the space usage. For most natural data structure problems, $m$ is only polynomial in the input size $n$ and $S \geq n$. This means that for most problems, the lower bounds degenerates to $\Omega(1)$.

This barrier was overcome in the seminal papers of \Patrascu and Thorup~\cite{PT:pred,patrascu10higher} where they introduced a refined reduction from communication complexity that pushed the barrier to lower bounds of $t = \Omega(\lg m/\lg(S m/n))$. Extending upon ideas of Panigrahy et al.\@~\cite{PTW:2010}, Larsen~\cite{Larsen:2012:focs} tweaked their cell sampling technique to give slightly higher lower bounds of $\Omega(\lg m/\lg(S/n))$. This remains the highest static lower bound to date.

\subsection{Our Results}

\paragraph{\Patrascu's Multiphase Problem.} 
In the Multiphase Problem, we have three phases. In Phase I, we receive $k$ subsets $X_1,\dots,X_k$ of a universe $[n]$ and must preprocess these into a data structure. In Phase II, we receive another set $Y \subseteq [n]$ and we are allowed to update our data structure based on this set $Y$. Finally, in Phase III, we receive an index $i \in [k]$ and the goal is to return whether $X_i \cap Y = \emptyset$. The three performance metrics of interest to us are the following: The \emph{space usage}, $S$, is defined as the number of memory cells of $w = \Omega(\lg k)$ bits used by the data structure produced in Phase I. The \emph{update time}, $t_u$, is the number of probes used in Phase II. The \emph{query time}, $t_q$, is the number of probes spend in Phase III.

As mentioned earlier, \Patrascu showed hardness results for the Multiphase Problem by a reduction from 3SUM. His reduction shows that for $k=\Theta(n^{2.5})$, it is 3SUM hard to design a word-RAM data structure for the Multiphase Problem that simultaneously spends $kn^{0.5-\Omega(1)}$ time in Phase I, $n^{1.5-\Omega(1)}$ time in Phase II and $n^{0.5-\Omega(1)}$ time in Phase III. Proving such polynomial lower bounds in the cell probe model is far out of reach. Nevertheless, we still find it extremely important to see what actually can be said unconditionally and try to understand the limitations of our techniques better.

In Section~\ref{sec:multi}, we introduce a new technique for proving strong \emph{threshold} lower bounds for dynamic data structures. We apply our technique to the Multiphase Problem and show the following: Any cell probe data structure for the Multiphase Problem with $w^4 \leq n \leq k$, using space $k n^{O(1)}$ cells of $w = \Omega(\lg k)$ bits and answering queries in $o(\lg k/\lg n)$ probes, must have $t_u = k^{1-o(1)}/w$. This lower bound holds even if the set $Y$ inserted in Phase II has size $O(\lg k/\lg n)$.

In the most natural case of $w = \Theta(\lg k)$, we can set $n = \lg^4 k$ and the lower bound says that any data structure for the Multiphase Problem with $\lg^4 k$-sized sets, which uses $k \lg^{O(1)} k$ words of space and supports queries in $o(\lg k/\lg \lg k)$ time, must have update time $k^{1-o(1)}$. And this applies even if $Y$ has size $O(\lg k/\lg \lg k)$. This lower bound has quite a remarkable statement: If we want to do anything better in Phase III than checking each element in $Y$ one at a time for inclusion in $X_i$, then Phase II has to compute a table of all the answers to all the $k$ possible queries. There is essentially no strategy in-between the two extremes.

The previous result that comes closest in spirit to our new lower bound is the threshold results of \Patrascu and Thorup~\cite{PT:stoc:2011}, showing that any data structure for dynamic connectivity in undirected graphs with $n$ nodes, having update time $t_u = o(\lg n)$, must have query time $t_q = n^{1-o(1)}$. Thus their lower bound is essentially the opposite way around. 

Since our lower bound is proved for the Multiphase Problem, we immediately get a similar lower bound for a number of problems, simply by reusing the previous conditional hardness reductions. We mention two examples from~\cite{Patrascu:2010:polylower} here: For dynamic connectivity in directed graphs with $n$ nodes and $m = n\lg^{O(1)} n$ edges, any data structure using $m \lg^{O(1)}m$ space and supporting connectivity queries in $o(\lg n/ \lg \lg n)$ time, must have update time $n^{1-o(1)}$. For dynamic shortest paths in undirected graphs with $n$ nodes and $m = n\lg^{O(1)} n$ edges, any data structure using $m \lg^{O(1)}m$ space and supporting distance queries in $o(\lg n/\lg \lg n)$ time, must have update time $n^{1-o(1)}$. Both lower bounds hold even if one node in the connectivity/distance query is a fixed source node (common to all queries).

\paragraph{Online matrix-vector multiplication.}
Given an $n \times n$ matrix $M$ with coefficients from a finite field $\F$, preprocess $M$ into a data structure, such that when given a query vector $v \in \F^n$, we can quickly compute $Mv$.

In Section~\ref{sec:online}, we show a lower bound of
$$t = \Omega{\left(\min\left\{\frac{n \lg |\F|}{\lg\left(\frac{Sw}{n^2 \lg |\F|}\right)}, \frac{n^2\lg|\F|}{w} \right\}\right)}$$
cell probes to compute $Mv$ for a query vector $v \in \F^n$. This holds even if the data structure is allowed to err with probability $1-1/|\F|^{n/4}$ on average over all pairs of a matrix $M$ and a query vector $v$. This is the first lower bound of this type which applies even under such extreme probability of error. 

For the natural range of parameters $|\F| = n^{\Omega(1)}$ and $w = \Theta(\lg |\F|)$, the lower bound simplifies to $t = \Omega(\min\{n \lg|\F|/\lg(S/n), n^2\})$ and is the strongest current techniques can show for a static problem with $|\F|^n$ queries. For linear space, this is $t=\Omega(\min\{n \lg |\F|, n^2\})$. As the size of the field $\F$ tends to $2^n$, the lower bound says that any data structure with near-linear space has to ``read'' the entire matrix to compute $Mv$, even if allowed to err with overwhelming probability. While this might sound odd at first, note that it also means that any data structure that doesn't explicitly try to exploit that it is dealing with a small field is doomed to use $n^2$ time to compute $Mv$.

Frandsen et al.\@~\cite{FHM:2001} also proved lower bounds for online matrix-vector multiplication, where the first term in the min-expression above is replaced by $n \lg|\F|/\lg S$. Their lower bound thus also shows that as the field size grows, the trivial solution is the only option. Comparing their lower bound to ours, we see that for linear space, our lower bound is a factor $\lg n$ stronger. Furthermore, their lower bound holds only for deterministic data structures, whereas ours allow an only exponentially small probability of returning the correct answer. 

\paragraph{Dynamic online matrix-vector multiplication.}
 Maintain an $n \times n$ matrix $M$ with coefficients from a finite field $\F$ under
 \begin{itemize}
  \item updates of the form $M_{i,j} \gets x$ for a row index $i$, column index $j$ and an $x \in \F$;
  \item matrix-vector queries.  Given an $n$-vector $v$ return the product $Mv$. 
 \end{itemize}  
  
In Section~\ref{sec:dynamic}, we prove that any cell probe data structure for the dynamic online matrix-vector multiplication problem on an $n \times n$ matrix $M$, with $w$ bit cells and worst case update time $t_u$, must use 
$$t_q = \Omega{\left(\min\left\{\frac{n \lg |\F| \lg(n/w)}{\lg^2\left(\frac{t_uw}{\lg |\F|}\right)}, \frac{n^2\lg|\F|}{w} \right\}\right)}$$
cell probes to compute $Mv$ for a query vector $v \in \F^n$. This holds if the data structure errs with probability no more than $1/3$ when answering any query vector $v$ after any sequence of $n^2$ updates. The lower bound we prove for dynamic online matrix-vector multiplication equals the highest that have ever been achieved for any dynamic data structure problem (in fact it is slightly stronger than any previous bound for update time $t_u = O(\lg|\F|/w)$, making it the strongest to date). It is also only the second example, after~\cite{Larsen:2012:focs}, of such a lower bound that holds under constant probability of error.

Given that progress on proving these $(\lg m \lg n)$-type dynamic lower bounds has been very slow, we find it an important contribution in itself to give a new lower bound of this form. We hope that the proof eventually will inspire new ways of proving lower bounds and will push the barriers further. In particular, one of the biggest problems with the current lower bound technique of Larsen~\cite{Larsen:2012} is that it can only be applied for problems where the answer to a query carries more information (bits/entropy) than it takes to describe a query. This in particular implies that the technique cannot be applied to decision problems. Our proof of the above lower bound makes some progress on this frontier. In the proof, we eventually end up with a collection of queries whose answers ``reveal'' only a small constant fraction of the ``information'' needed to describe them. We elegantly circumvent the limitations of the lower bound technique by using a randomized encoding argument that allows us to save a constant fraction in the ``description size'' of the queries. We refer the reader to the proof itself for the details.

\section{Threshold Bounds for The Multiphase Problem}
\label{sec:multi}
We prove our lower bound for the Multiphase Problem in the cell probe model. The lower bound we prove is the following
\begin{theorem}
\label{thm:multiphase}
Any cell probe data structure for the Multiphase Problem on $k$ sets from a universe $[n]$, where $w^4 \leq n \leq k$, using $k n^{O(1)}$ cells of $w = \Omega(\lg k)$ bits of space and answering queries in $t_q = o(\lg k/\lg n)$ probes, must have $t_u = k^{1-o(1)}/w$. This holds even if the update set in Phase II contains $O(\lg k/\lg n)$ elements.
\end{theorem}
We prove Theorem~\ref{thm:multiphase} by a reduction from a variant of the communication game \emph{Lop-sided Set Disjointness}, or LSD for short. In LSD, we have two players Alice and Bob. Alice and Bob receive subsets $V$ and $W$ of a universe $[U]$ and must determine whether $V \cap W = \emptyset$ while minimizing their communication. The term Lop-sided stems from Alice's set having size $N$, where $NB = U$ for some value $B >1$. As mention, we use a variant of LSD known as Blocked-LSD~\cite{Patrascu:2008:unifyingcellprobe}. In Blocked-LSD, the universe is the Cartesian product $[N] \times [B]$. Bob receives a subset $W$ of $[N] \times [B]$, which may be of arbitrary size. Alice's set $V$ satisfies that for all $j \in [N]$, there is exactly one $b_j \in [B]$ such that $(j,b_j) \in V$, i.e. $V$ has the form $\{(0,b_0),\dots,(N-1,b_{N-1})\}$. P{\v a}tra{\c s}cu proved the following lower bound for Blocked-LSD:
\begin{theorem}[P{\v a}tra{\c s}cu~\cite{Patrascu:2008:unifyingcellprobe}]
\label{thm:blockedlsd}
Fix $\delta > 0$. In any deterministic protocol for Blocked-LSD, either Alice sends $\delta N \lg B$ bits or Bob sends $NB^{1-O(\delta)}$ bits.
\end{theorem}
In our reduction, we will need the following lemma:
\begin{lemma}
\label{lem:hashsubset}
Consider a communication game in which Bob receives a set $B \subseteq [2^w]$ of size $S$ and Alice receives a set $A \subseteq B$ of size $k$. There is a deterministic protocol in which Alice sends $O(k \lg(S/k))$ bits, Bob sends $O(kw)$ bits, and after communicating, Bob knows Alice's set $A$.
\end{lemma}
The proof of Lemma~\ref{lem:hashsubset} is based on a simple application of hashing and is given in Section~\ref{sec:hashsubset}. We note that a similar trick has been used by Miltersen~\cite{Miltersen:1994}, but only for $k=1$. His proof thus ``costs'' $\lg S$ bits in Alice's communication per element in $A$, whereas we shave this down to $\lg(S/k)$ bits. We are now ready to give the reduction from Blocked-LSD to the Multiphase Problem.
\begin{proof}[Proof (of Theorem~\ref{thm:multiphase})]
Assume we have a data structure $\D$ for the Multiphase Problem with $k$ sets from the universe $[n]$, where $w^4 \leq n \leq k$.  Let $S$ be the space usage of $\D$ in number of cells of $w = \Omega(\lg k)$ bits each. Let $t_q$ be its query time and $t_u$ its update time. We assume $S = kn^{O(1)}$ and $t_q = o(\lg k/\lg n)$ and show this implies $t_u = k^{1-o(1)}$. Note that for this setting of parameters, we have $n = k^{o(1)}$ since otherwise it is impossible to have $t_q = o(\lg k/\lg n)$.

Define $\ell = \sqrt{t_q \lg k/\lg n}$. Since we assumed $t_q = o(\lg k/\lg n)$, we have $\ell = o(\lg k/\lg n)$ and $\ell = \omega(t_q)$.
 We use $\D$ to give an efficient communication protocol for Blocked-LSD on the universe $[k \ell] \times [n/\ell]$. For this setting of parameters, Theorem~\ref{thm:blockedlsd} says that either Alice sends $\Omega(k \ell \lg(n/\ell)) = \omega(k t_q \lg n)$ bits, or Bob sends $k \ell (n/\ell)^{3/4} \geq k \ell (w^3/\ell) = \omega(k w^2) $ bits.

Alice receives $V$ and Bob receives $W$, both subsets of $[k \ell] \times [n/\ell]$. Alice's set $V$ satisfies that for all $j \in [k \ell]$, there is exactly one $b_j \in [n/\ell]$ such that $(j,b_j) \in V$. Alice and Bob now conceptually partition $[k\ell]$ into $k$ consecutive groups $G_1,\dots,G_{k}$ of $\ell$ elements each, i.e. the first group is $G_1 = \{0,\dots,\ell-1\}$, the second is $G_2 = \{\ell,\dots,2\ell-1\}$ etc. For $i=1,\dots,k$ we let $V_i$ denote the subset of pairs $(j,b_j) \in V$ for which $j \in G_i$. Similarly we let $W_i$ denote the subset of pairs $(j,h) \in W$ for which $j \in G_i$. Observe that $|V_i| = \ell$ for each $i$. There is no size bound on $W_i$ other than the trivial bound $|W_i| \leq \ell(n/\ell) = n$.

Alice now interprets each of the subsets $V_i \subseteq \{i\ell,\dots,(i+1)\ell-1\} \times [n/\ell]$ as an $\ell$-sized subset of the universe $[n]$, denoted $Y_i$. This is done by mapping a pair $(j,b_j) \in V_i$ to the element $(j\mod \ell)(n/\ell) + b_j$. Bob similarly interprets each of his subsets $W_i \subseteq \{i\ell,\dots,(i+1)\ell-1\} \times [n/\ell]$ as a subset of the universe $[n]$, denoted $X_i$. He also does this by mapping a pair $(j,h) \in W_i$ to the element $(j\mod \ell)(n/\ell) + h$. The crucial property of this reduction is that $V \cap W = \emptyset$ if and only if $X_i \cap Y_i = \emptyset$ for all $i=1,\dots,k$. The goal now is for Alice and Bob to use $\D$ to test whether $X_i \cap Y_i = \emptyset$ for all $i=1,\dots,k$ and thereby determine whether $V \cap W = \emptyset$. This is done using the following protocol:
\begin{enumerate}
\item Bob starts by running Phase I of the Multiphase Problem on $\D$ with the sets $X_1,\dots,X_k$ as input. This creates a data structure using only $S = k n^{O(1)}$ memory cells. 
 Note that Bob does not communicate with Alice in this step and thus the constructed data structure is only known to Bob.
\item Alice now iterates through all $\ell$-sized subsets of $[n]$. For each such subset $Y$, she runs Phase II of the Multiphase Problem on the data structure held by Bob with $Y$ as input. This is done as follows: For a subset $Y$, Alice first initializes an empty set of cells $C(X_1,\dots,X_k, Y)$. She then starts running the update algorithm of $\D$ with $Y$ as input. This either requests a memory cell or overwrites the contents of a memory cell. In the latter case, Alice stores the overwritten cell in $C(X_1,\dots,X_k,Y)$, including its address and its new contents. In the first case, Alice checks whether the requested cell is in $C(X_1,\dots,X_k,Y)$. If so, she has the contents herself and can continue running the update algorithm. Otherwise, she asks Bob for the contents of the cell by sending him $w$ bits specifying the cell's address. Bob then replies with its $w$ bits of contents. When this terminates, each of the cell sets $C(X_1,\dots,X_k,Y)$ held by Alice stores the contents and addresses of every cell that is updated if running Phase II on $\D$ with $Y$ as input, after having run Phase I on $\D$ with the sets $X_1,\dots,X_k$ as input. Since $\D$ has update time $t_u$, Alice and Bob both send no more than $t_u w$ bits for each $\ell$-sized subset of $[n]$. Since we chose $\ell = o(\lg k/\lg n)$, this is no more than $k^{o(1)}t_u w$ bits in total. Note that Bob performs no other actions in this step than to reply to Alice with the contents of the requested cells (with the contents right after processing $X_1,\dots,X_k$ in Phase I).
\item Alice now runs the Phase III query algorithm of $\D$ for every possible query $i \in [k]$ \emph{in parallel}. The execution for a query index $i$ will be run as if the updates $Y_i$ had been performed in Phase II. The query $i$ will thus return whether $X_i \cap Y_i = \emptyset$. More formally, Alice does as follows: For $t=1,\dots,t_q$ in turn, Alice will simulate the $t$'th probe of $\D$ for every query $i \in [k]$. She will do this so that the execution is identical to having run the updates  $Y_i$ in Phase II. For the $t$'th probe, the query algorithm of $\D$ requests a memory cell $c_{t,i}$ for each $i$. For the cell $c_{t,i}$, she checks whether that cell is contained in $C(X_1,\dots,X_k, Y_i)$. If so, she has the contents of the cell as if update $Y_i$ was performed in Phase II and she can continue to the next probe for that $i$ without communicating with Bob. If not, she knows that the contents of $c_{t,i}$ was not changed when performing the updates $Y_i$ in Phase II. She then adds the address of $c_{t,i}$ to a set of addresses $Z_t$. The set $Z_t$ thus holds the addresses of all cells needed to execute the $t$'th probe for each query $i \in [k]$, and Alice needs the contents of these cells as they were right after Phase I. Alice will now ask Bob for the contents of all cells in $Z_t$. The point of collecting the cells needed in one set $Z_t$, rather than asking for them one at a time, is to save on the communication, i.e. Alice wants to send less than $w$ bits (the address) to Bob per cell in $Z_t$. This is done by invoking Lemma~\ref{lem:hashsubset}, with the $B$ in Lemma~\ref{lem:hashsubset} being the addresses of all cells written to in Phase I on input $X_1,\dots,X_k$ and $A$ is the set $Z_t$. After using Lemma~\ref{lem:hashsubset}, Bob knows $Z_t$ and sends the contents of all cells in $Z_t$ to Alice. By Lemma~\ref{lem:hashsubset}, Bob will send $O(kw)$ bits and Alice will send $O(k \lg(S/k)) = O(k \lg n)$ bits. Alice can now continue with probe $t+1$ and eventually the data structures determines whether $X_i \cap Y_i = \emptyset$ for each $i$. Since $X_i \cap Y_i = \emptyset$ for each $i$ iff $V \cap W = \emptyset$, this completes the description of the protocol.
\end{enumerate}
We have thus given a protocol for Blocked-LSD on $[k \ell] \times [n/\ell]$ in which Alice sends $k^{o(1)}t_u w + O(t_q k \lg n)$ bits and Bob sends $k^{o(1)}t_u w + O(t_q k w) = k^{o(1)}t_u w + o(kw^2)$ bits. But the lower bound says that either Alice must send $\omega(t_q k \lg n)$ bits or Bob must send $\omega(kw^2)$ bits. This implies $t_u w k^{o(1)} = \omega(k) \Rightarrow t_u = k^{1-o(1)}/w$.
\end{proof}

\subsection{Communicating a Subset (Proof of Lemma~\ref{lem:hashsubset})}
\label{sec:hashsubset}
Let $B \subseteq [2^w]$ with $|B| = S$ and let $A \subseteq B$ with $|A| = k$. Bob receives $B$ and Alice receives $A$. Consider the $2/2^M$-universal hash function $h_a(x) = \lfloor (a x \mod 2^w)/2^{w-M}\rfloor$ of Dietzfelbinger et al.~\cite{Dietzfelbinger}, where $a$ is a uniform random odd integer less than $2^w$. By $2/2^M$ universal we mean that for any two distinct $x,y \in [2^w]$, we have $\Pr_a[h_a(x) = h_a(y)] \leq 2/2^M$ (note there are $2^M$ possible values $h_a(x)$ can take). Letting $M = \lceil \lg S \rceil$, we have for any two distinct $b_1,b_2 \in B$ that $\Pr_a[h_a(b_1) = h_a(b_2)] \leq 2/S$. The expected number of distinct pairs $(b_1,b_2) \in B$ for which $h_a(b_1) = h_a(b_2)$ is at most $2S$. Thus there must exist an odd integer $a^* \in [2^w]$ such that the number of pairs $(b_1,b_2) \in B$ where $h_{a^*}(b_1) = h_{a^*}(b_2)$ is at most $2S$. Bob starts by sending Alice such an odd integer $a^*$, costing $w$ bits of communication from Bob. Alice now computes $h_{a^*}(A) \subseteq [2^M]$ and sends $h_{a^*}(A)$ to Bob by specifying it as a subset of $[2^M]$. Since $|h_{a^*}(A)|\leq k$ and $2^M \leq 2S$, this costs at most $\lg \binom{2S}{k} = O(k \lg(S/k))$ bits. For each $i$ in $h_{a^*}(A)$, Bob computes the set $B_i$ consisting of all elements in $b \in B$ such that $h_{a^*}(b) = i$. Since the total number pairs $b_1,b_2 \in B$ with $h_{a^*}(b_1)=h_{a^*}(b_2)$ is no more than $2S$, we have $\sum_{i \in h_{a^*}(A)} |B_i|^2 \leq 2S$. For each $i \in h_{a^*}(A)$, Bob now picks $M_i = \lceil \lg \left(8|B_i|^2\right) \rceil$ and finds an odd integer $a_i^* \in [2^{M_i}]$ such that for all $b_1,b_2 \in B_i$, we have $h_{a_i^*}(b_1) \neq h_{a_i^*}(b_2)$. Bob sends all these $a^*_i$'s to Alice, costing at most $O(|h_{a^*}(A)|w) = O(kw)$ bits. Finally Alice computes for each $i \in h_{a^*}(A)$ the set $A_i$ of elements $a \in A$ such that $h_{a^*}(a)=i$. For each $A_i$, Alice computes $h_{a^*_i}(A_i) \subseteq [2^{M_i}]$. Since $\sum_{i \in h_{a^*}(A)} 2^{M_i} = O(S)$, Alice can now send $h_{a^*_i}(A_i)$ to Bob for every $i$ with a total communication of at most $\lg \binom{O(S)}{k} = O(k \lg(S/k))$ bits. Since $A \subseteq B$ and $h_{a^*_i}(b_1) \neq h_{a^*_i}(b_2)$ for any two $b_1,b_2 \in B_i$, Bob has learned $A$.

In the protocol above, Bob sends $O(kw)$ bits and Alice sends $O(k \lg(S/k))$ bits. Note that the protocol is deterministic, the randomness of the hash functions is only used to argue that there exists a choice of $a^*$ and $a_i^*$'s. We have thus proved Lemma~\ref{lem:hashsubset}.
\section{Online Matrix-Vector Multiplication}
\label{sec:online}
In this section, we consider the online matrix-vector multiplication problem: Given an $n \times n$ matrix $M$ with coefficients from a finite field $\F$, preprocess $M$ into a data structure, such that when given a query vector $v \in \F^n$, we can quickly compute $Mv$. We consider the problem in the cell probe model with $w$ bit cells where $w$ is assumes to be at least $\lg n$ and at least $\lg |\F|$. Our lower bound is as follows:

\begin{theorem}
\label{thm:staticonline}
Any cell probe data structure for the online matrix-vector multiplication problem, using $S$ cells of $w$ bits of space to store an $n \times n$ matrix with coefficients from a finite field $\F$, must use 
$$t = \Omega\left(\min\left\{\frac{n \lg |\F|}{\lg\left(\frac{Sw}{n^2 \lg |\F|}\right)}, \frac{n^2\lg|\F|}{w} \right\}\right)$$
cell probes to compute $Mv$ for a query vector $v \in \F^n$. This holds even if the data structure is allowed to err with probability $1-1/|\F|^{n/4}$ on average over all pairs of a matrix $M$ and a query vector $v$.
\end{theorem}

For the natural range of parameters $|\F| = n^{\Omega(1)}$ and $w = \Theta(\lg |\F|)$, the lower bound simplifies to $t = \Omega(\min\{n \lg|\F|/\lg(S/n), n^2\})$. For linear space, this is $t=\Omega(\min\{n \lg |\F|, n^2\})$. As the size of the field $\F$ tends to $2^n$, the lower bound says that any data structure with near-linear space has to ``read'' the entire matrix to compute $Mv$, even if allowed to err with overwhelming probability.

We give the proof in the following. The proof is based on an encoding argument.
\paragraph{Encoding Argument.}
  Consider a randomized data structure $\D$ for online matrix-vector multiplication using $S$ cells of $w$ bits of space. Assume the data structure answers queries in $t$ probes with error probability $1-1/|\F|^{n/4}$ on average over all pairs of an input matrix $M$ and query vector $v$. Now consider the following hard distribution: The input matrix is a uniform random matrix $M$ in $\F^{n \times n}$ and the query to be answered after preprocessing is a uniform random $v \in \F^n$. By fixing the random coins of $\D$, there exists a deterministic data structure $\D^*$ with space $S$ cells of $w$ bits, query time $t$ and error probability $1-1/|\F|^{n/4}$ over the hard distribution. Using Markov's inequality, we conclude that there must be a family of matrices $\Fam \subseteq \F^{n \times n}$, with $$|\Fam| \geq |\F|^{n^2} \left(1-\frac{1-\frac{1}{|\F|^{n/4}}}{1-\frac{1}{|\F|^{n/2}}}\right) =|\F|^{n^2} \left(\frac{\frac{1}{|\F|^{n/4}}-\frac{1}{|\F|^{n/2}}}{1-\frac{1}{|\F|^{n/2}}}\right) \geq |\F|^{n^2-n/2},$$
such that for every matrix $M \in \Fam$, $D^*$ answers at least $|\F|^n/|\F|^{n/2} = |\F|^{n/2}$ of the possible query vectors $v$ correctly after having preprocessed $M$. To derive the lower bound, we show that $\D^*$ can be used to efficiently encode every matrix $M \in \Fam$ into a bit string with length depending on $t, S, w$ and $|\F|$. If every $M \in \Fam$ can be uniquely recovered from these bit strings, we know that at least one of the bit strings must have length $\lg |\Fam| \geq (n^2-n/2)\lg|\F|$ resulting in a lower bound trade-off for $t,S,w$ and $|\F|$.

To encode a matrix $M \in \Fam$, we do as follows:
\begin{enumerate}
\item Construct $\D^*$ on $M$. This gives a memory representation consisting of $S$ cells of $w$ bits. Now iterate over all vectors $v \in \F^n$ and collect the subset $V$ consisting of those vectors $v$ for which $\D^*$ does not err when answering $v$ after having preprocessed $M$. Since $M \in \Fam$, we know $|V| \geq |\F|^{n/2}$.
\item Interpret every vector $v \in \F^n$ as an integer $f(v)$ in the range $[\F^n] = \{0,\dots,|\F|^n-1\}$ in the natural way $f(v) = \sum_{i=1}^n v(i) |\F|^{i-1}$. Consider the random hash function $h_a : [\F^n] \to [\F^{n/8}]$ with $h(x) = (x+a) \mod |\F|^{n/8}$ for a uniform random $a \in [\F^{n/8}]$. Let $W^0_a$ denote the set of all vectors $v \in \F^n$ for which $h_a(f(v)) = 0$. We always have $|W^0_a|=|\F|^{7n/8}$. Furthermore, $\E_a[|W^0_a \cap V|] = |V|/|\F|^{n/8} \geq |\F|^{3n/8}$. Hence there exists a choice of $a \in [\F^{n/8}]$ such that $|W^0_a \cap V| \geq |\F|^{3n/8}$. The first part of the encoding is such a value $a^* \in [\F^{n/8}]$, costing $3n\lg|\F|/8$ bits.
\item Having chosen $a^*$, we now consider every set $C$ of $\Delta = n^2\lg|\F|/(1024w)$ memory cells in the data structure. For each such set $C$, let $Q_C$ denote the set of query vectors $v \in \F^n$ for which $\D^*$ probes only cells in $C$ when answering $v$ after having preprocessed $M$. We let $C^*$ be the the set of $\Delta$ memory cells for which $|Q_{C^*} \cap W^0_{a^*} \cap V|$ is largest. We then write down the addresses and contents of cells in $C^*$. This costs no more than $\Delta(w + \lg S) \leq 2\Delta w = n^2\lg|\F|/512$ bits.
\item We now consider the set of query vectors $V^* = Q_{C^*} \cap W^0_{a^*} \cap V$. Since any $k$-dimensional subspace of $\F^n$ contains at most $|\F|^k$ vectors, we know that $\dim(\vspan(V^*)) \geq \lg_{|\F|} |V^*|$. We can thus find a set of $\lg|V^*|/\lg|\F|$ linearly independent vectors in $V^*$. We write down such a set of vectors $U$. Since $U \subseteq W^0_{a^*}$, we can specify $U$ as indices into $W^0_{a^*}$, costing only $(\lg|V^*|/\lg|\F|)(7n\lg|\F|/8) = 7n\lg|V^*|/8$ bits.
\item Finally we initialize a set of vectors $X = \emptyset$ and iterate through all vectors in $\F^n$ in lexicographic order. For such vector $x$, we check if $x \in \vspan(U \cup X)$. If so, we continue to the next vector. If not, we add $x$ to $X$. This terminates with $\dim(\vspan(U \cup X))=|U| + |X|=n$. In the last step of our encoding procedure, we examine each row vector $m_i$ of $M$ in turn. For the $i$'th row vector, we compute the inner product $\langle m_i , x\rangle$ over $\F$ for every $x \in X$. We write down each of these $|X|$ inner products for a total of $n|X|\lg|\F|$ bits. This concludes the description of the encoding procedure.
\end{enumerate}
Before presenting the decoding procedure, we make a few remarks regarding the ideas in the above encoding procedure. Intuitively each query in $U$ can be answered solely from the contents of $C^*$. Furthermore, the query vectors in $U$ are linearly independent and thus in total reveal $|U|\lg|\F|$ bits of information about each of the $n$ rows of $M$. The hashing trick in steps 2-3 ensure that the vectors in $U$ can be described using only $\lg |W^0_{a^*}| = 7n\lg|\F|/8$ bits each. Thus each vector reveals $n\lg|\F|/8$ more bits of information about $M$ than it costs to describe. Thus $U$ will have to be small, leading to a space time trade off.

We now show how $M$ can be recovered from the encoding produced above. The decoding procedure is as follows:
\begin{enumerate}
\item From the bits written during step 2. and 3. of the encoding procedure, we recover $a^*$ and $C^*$. From $a^*$ we also obtain $W^0_{a^*}$.
\item Now that $W^0_{a^*}$ has been recovered, we use the bits written in step 4. of the encoding procedure to recover the set of vectors $U$. We now run the query algorithm of $D^*$ for every $v \in U$. Since $U \subseteq Q_{C^*}$, the query algorithm only probes cells in $C^*$ when answering these queries. Since we have the addresses and contents of all cells in $C^*$, we thus obtain $Mv$ for every $v \in U$, i.e. we know $\langle m_i , v\rangle$ for every row vector $m_i$ and every $v \in U$.
\item Finally we initialize an empty set of vectors $X = \emptyset$ and iterate through all vectors $x \in \F^n$ in lexicographic order. For each vector $x$, we check if $x \in \vspan(U \cup X)$. If so, we continue to the next vector. If not, we add $x$ to $X$ and continue. This recovers the exact same set of vectors $X$ as in step 5. of the encoding procedure. From the bits written during step 5. of the encoding procedure, we obtain $\langle m_i ,x \rangle$ for every $x \in X$. Since $\dim(\vspan(U \cup X))=n$ and we know $\langle m_i , u \rangle$ for every $u \in U \cup X$, this uniquely determines $m_i$ which completes the decoding procedure.
\end{enumerate}

\paragraph{Analysis.}
Above we argued that the above procedures allow us to encode and decode every matrix $M \in \Fam$ into a bit string. Thus there must be a matrix $M \in \Fam$ for which the bit string produced has length at least $\lg |\Fam| \geq (n^2-n/2)\lg|\F|$. But the encoding produced has length
$$
3n\lg|F|/8 + n^2\lg|\F|/512 + 7n \lg|V^*|/8 + n|X|\lg|\F|
$$
bits. Since $|U| = \lg|V^*|/\lg|\F|$, we have $|X| = n-|U| = n-\lg|V^*|/\lg|\F|$ and we conclude that we must have
\begin{eqnarray*}
(n^2-n/2)\lg|\F| &\leq& 3n\lg|\F|/8 + n^2\lg|\F|/512 + 7n\lg|V^*|/8 + n(n-\lg|V^*|/\lg|\F|)\lg|\F| \\
&=& 3n\lg|\F|/8 + n^2\lg|\F|/512  + n^2\lg|\F| - n\lg|V^*|/8
\end{eqnarray*}
This implies
\begin{eqnarray*}
n\lg|V^*|/8 &\leq& 7n\lg|\F|/8 + n^2\lg|\F|/512 \Rightarrow\\
\lg|V^*| &\leq& 7\lg|\F| + n\lg|\F|/64 \Rightarrow\\
\lg|V^*| &\leq& n\lg|\F|/32.
\end{eqnarray*}
Since $C^*$ was chosen such that $|V^*|$ was largest possible, we know by averaging that if the data structure has query time $t \leq \Delta/2$, then 
\begin{eqnarray*}
  |V^*| &\geq& \frac{|W^0_{a^*} \cap V| \binom{S-t}{\Delta-t}}{\binom{S}{\Delta}}\\
&=& \frac{|W^0_{a^*} \cap V| (S-t)!(S-\Delta)!\Delta!}{S!(S-\Delta)!(\Delta-t)!}\\
&=& \frac{|W^0_{a^*} \cap V| (S-t)!\Delta!}{S!(\Delta-t)!}\\
&\geq& \frac{|W^0_{a^*} \cap V| (\Delta-t)^t}{S^t}\\
&\geq& |W^0_{a^*} \cap V| \left(\frac{\Delta}{2S}\right)^t\\
&\geq& |\F|^{3n/8} \left(\frac{n^2 \lg|\F|}{2048 Sw}\right)^t
\end{eqnarray*}
Taking logs, we conclude that we must have
\begin{eqnarray*}
3n\lg|\F|/8 + t \lg\left(\frac{n^2 \lg|\F|}{2048 S w}\right) &\leq& n\lg|\F|/32 \Rightarrow\\
t \lg\left(\frac{2048 S w}{n^2 \lg|\F|}\right) &\geq& 11n \lg|\F|/32 \Rightarrow\\
t &=& \Omega\left(\frac{n \lg|\F|}{\lg\left(\frac{Sw}{n^2 \lg|\F|}\right)}\right).
\end{eqnarray*}
Since the above calculation needed $t \leq \Delta/2 = n^2 \lg|\F|/(2048w)$, we conclude that 
$$t = \Omega\left(\min\left\{ \frac{n \lg|\F|}{\lg\left(\frac{Sw}{n^2 \lg|\F|}\right)}, \frac{n^2 \lg|\F|}{w} \right\}\right).$$

\section{Dynamic Online Matrix-Vector Multiplication}
\label{sec:dynamic}
In this section, we prove a lower bound for the dynamic online matrix-vector multiplication problem: Maintain an $n \times n$ matrix $M$ with coefficients from a finite field $\F$, such that we can efficiently support entry updates of the form $m_{i,j} \gets x$ for a row index $i$, column index $j$ and an $x \in \F$. The matrix $M$ is initialized to the all $0$'s matrix, and at any time, we may ask a query $v \in \F^n$ and the data structure must return $Mv$. We prove the following lower bound for this problem:
\begin{theorem}
\label{thm:maindynamic}
Any cell probe data structure for the dynamic online matrix-vector multiplication problem on an $n \times n$ matrix $M$, with $w$ bit cells and worst case update time $t_u$, must use 
$$t_q = \Omega\left(\min\left\{\frac{n \lg |\F| \lg(n/w)}{\lg^2\left(\frac{t_uw}{\lg |\F|}\right)}, \frac{n^2\lg|\F|}{w} \right\}\right)$$
cell probes to compute $Mv$ for a query vector $v \in \F^n$. This holds if the data structure errs with probability no more than $1/3$ when answering any query vector after any sequence of $n^2$ updates.
\end{theorem}
To prove the theorem, we follow the general approach ventured in~\cite{Larsen:2012}. The first step is to define a hard distribution. 

\paragraph{Hard Distribution.}
For the dynamic online matrix-vector multiplication problem, our hard distribution is as follows: Let $(i_{n^2},j_{n^2},x_{n^2}),(i_{n^2-1},j_{n^2-1},x_{n^2-1}),\dots,(i_{1},j_{1},x_{1})$ be a sequence of updates to the matrix $M$. The triple $(i_{n^2},j_{n^2},x_{n^2})$ is the first update and $(i_{1},j_{1},x_{1})$ is the last update. A triple $(i_k,j_k,x_k)$ corresponds to the update operation $M_{i_k,j_k} \gets x_k$. The values $x_k$ are uniform random and independent in $\F$. The sequence of row and column indices $i_k,j_k$ is some fixed sequence of \emph{well-spread} indices, where well-spread is defined as follows:
\begin{definition}
\label{def:spread}
A sequence of row and column indices $(i_{n^2},j_{n^2}),\dots,(i_1,j_1)$ is well-spread if: 
\begin{itemize}
\item All pairs $(i_k,j_k)$ are distinct.
\item For every index $n^{4/3} \leq r \leq n^2$ and every set of $n/2$ row indices $S \subseteq \{1,\dots,n\}$, there exists a subset $S^* \subseteq S$ with $|S^*| \leq 8n^2/r$, such that $\left| \bigcup_{k \leq r : i_k \in S^*} \{j_k\}\right| \geq n/4$.
\end{itemize}
\end{definition}
Note that our hard distribution only needs that the sequence of update indices is well-spread. We do not care about the particular indices in the sequence. A well-spread sequence of indices basically guarantees that in every big enough set of rows (size at least $n/2$), there exists a small subset of rows ($S^*$), such that the indices updated in these rows ``cover'' at least $n/4$ columns. This must be true even if considering only the last $r$ updates for any $n^{4/3} \leq r \leq n^{2}$. The following lemma shows that such a well-spread sequence indeed exists:
\begin{lemma}
\label{lem:existspread}
There exists a well-spread sequence $(i_{n^2},j_{n^2}),\dots,(i_1,j_1)$ of row and column indices.
\end{lemma}
The proof of the lemma is a rather straight forward counting argument. We thus defer it to Section~\ref{sec:spread}.

Following the sequence of updates $(i_{n^2},j_{n^2},x_{n^2}),(i_{n^2-1},j_{n^2-1},x_{n^2-1}),\dots,(i_{1},j_{1},x_{1})$, we ask a uniform random query $v \in \F^n$. This concludes the description of our hard distribution.

By fixing the random coins, a randomized data structure $\D$ for dynamic online matrix-vector multiplication, with $w$ bit cells, worst case update time $t_u$, query time $t_q$ and error probability at most $1/3$ on any sequence of $n^2$ updates followed by a query, yields a deterministic data structure $\D^*$ with $w$ bit cells, worst case update time $t_u$, query time $t_q$ and error probability $1/3$ over the hard distribution. We thus continue by proving a lower bound for such a deterministic data structures.

\paragraph{Chronogram Approach.}
Following~\cite{Larsen:2012}, we partition the random updates  
$(i_{n^2},j_{n^2},x_{n^2}),\dots,(i_{1},j_{1},x_{1})$
into \emph{epochs} of roughly $\beta^\ell$ updates, where $\ell = 1,\dots,\lg_\beta n^2$ and $\beta > 2$ is a parameter to be fixed later. The $\ell$'th epoch consists of the $\beta^\ell - \beta^{\ell-1}$ updates $(i_{\beta^\ell-1},j_{\beta^\ell-1}, x_{\beta^\ell-1}),\dots,(i_{\beta^{\ell-1}},j_{\beta^{\ell-1}},x_{\beta^{\ell-1}})$. At the end of epoch $1$, the uniform random query $v \in \F^n$ is asked. 

When a deterministic data structure $\D^*$ processes the (random) sequence of updates 
$$\Pi = (i_{n^2},j_{n^2},x_{n^2}),\dots,(i_{1},j_{1},x_{1}),$$
we say that a memory cell \emph{belongs to} epoch $\ell$ if that memory cell's contents where last updated while processing the updates of epoch $\ell$, i.e. it \emph{was} updated during epoch $\ell$ and it was \emph{not} updated during epochs $\ell-1,\dots,1$. We let $C_\ell(\Pi)$ denote the set of memory cells belonging to epoch $\ell$ after processing $\Pi$. If $\D^*$ has worst case update time $t_u$, we have $|C_\ell(\Pi)| \leq \beta^{\ell}t_u$. We also define the set of probed cells $P(\Pi,v)$ as the set of memory cell probed when $\D^*$ answers the (random) query $v$ after processing the updates $\Pi$. With these definitions, the main technical challenge is to prove the following:
\begin{lemma}
\label{lem:epoch}
If $\D^*$ is a deterministic data structure for dynamic online matrix-vector multiplication, with $w$ bit cells, worst case update time $t_u$ and error probability $1/3$ over the hard distribution, then for all epochs $(4/6)\lg_\beta n^2 \leq \ell \leq \lg_\beta n^2$, we have
$$
\E_{\Pi,v}\left[|C_\ell(\Pi) \cap P(\Pi,v)|\right] = \Omega\left(\min\left\{\frac{n \lg |\F|}{\lg\left(\frac{t_uw}{\lg |\F|}\right)}  ,\frac{\beta^\ell \lg|\F|}{w}\right\}\right).
$$
assuming $\beta = 1024 t_uw/\lg|\F| $.
\end{lemma}
Before proving Lemma~\ref{lem:epoch}, we show that it implies Theorem~\ref{thm:maindynamic}. By the disjointness of the cell sets $C_{\lg_\beta n^3}(\Pi),\dots,C_1(\Pi)$ we always have $|P(\Pi,v)| \geq \sum_{\ell = 1}^{\lg_\beta n^2} |P(\Pi, v) \cap C_{\ell}(\Pi)|$. Thus by linearity of expectation we get
$$
\E_{\Pi,v}[|P(\Pi,v)|] \geq \sum_{\ell = 1}^{\lg_\beta n^2} \E_{\Pi,v}[|P(\Pi, v) \cap C_{\ell}(\Pi)|].
$$
By Lemma~\ref{lem:epoch}, this sum is at least
\begin{eqnarray*}
\E_{\Pi,v}[|P(\Pi,v)|] &\geq& \Omega\left(\sum_{\ell = (4/6)\lg_\beta n^2}^{\lg_\beta n^2} \min\left\{\frac{n \lg |\F|}{\lg\left(\frac{t_uw}{\lg |\F|}\right)}  ,\frac{\beta^\ell \lg|\F|}{w}\right\}\right).
\end{eqnarray*}
If $nw/\lg(t_u w/\lg|\F|) \geq n^2$, we get a lower bound of $\Omega(n^2 \lg|\F|/w)$ from Lemma~\ref{lem:epoch} applied only to epoch $\lg_\beta n^2$. If $nw/\lg(t_u w/\lg|\F|) \leq n^{3/2}$, the first term in the min expression is smallest for every index $\ell$ in the sum and we get a lower bound of $\Omega(n \lg|\F| \lg_\beta n/\lg(t_u w/\lg|\F|))$. If $n^{3/2} \leq nw/\lg(t_u w/\lg|\F|) \leq n^2$, there are $\lg_\beta n^2 - \lg_\beta (nw/\lg(t_u w/\lg|\F|)) \geq \lg_\beta(n/w)$ terms where the first term in the min expression is smallest, giving a lower bound of $\Omega(n \lg|\F| \lg_\beta (n/w)/\lg(t_u w/\lg|\F|))$. Since $\beta = 1024 t_u w/\lg|\F|$, this proves Theorem~\ref{thm:maindynamic}.

The next section is devoted to proving Lemma~\ref{lem:epoch}.
\subsection{Probes to Epoch $\ell$ (Proof of Lemma~\ref{lem:epoch})}
In this section we prove Lemma~\ref{lem:epoch}. Let $\D^*$ be a deterministic data structure for dynamic online matrix-vector multiplication with $w$ bit cells, worst case update time $t_u$ and error probability $1/3$ over the hard distribution. Let $\ell$ be an epoch satisfying $(4/6)\lg_\beta n^2 \leq \ell \leq \lg_\beta n^2$. Our goal is to prove that $$
\E_{\Pi, v}\left[|C_\ell(\Pi) \cap P(\Pi,v)|\right] = \Omega\left(\min\left\{\frac{n \lg |\F|}{\lg\left(\frac{t_uw}{\lg |\F|}\right)}  ,\frac{\beta^\ell \lg|\F|}{w}\right\}\right)
$$
assuming $\beta = 1024 t_uw/\lg |\F|$. Our proof is based on an encoding argument. More specifically, we assume for contradiction that
\begin{equation}
\label{eqn:assump}
\E_{\Pi, v}\left[|C_\ell(\Pi) \cap P(\Pi,v)|\right] = o\left(\min\left\{\frac{n \lg |\F|}{\lg\left(\frac{t_uw}{\lg |\F|}\right)}  ,\frac{\beta^\ell \lg|\F|}{w}\right\}\right)
\end{equation}
and use this assumption to encode the random updates of epochs $\ell,\dots,1$ in less than $H(x_{\beta^{\ell}-1} \cdots x_1) = \beta^\ell \lg |\F|$ bits in expectation. Here $H(\cdot)$ denotes binary Shannon entropy. By Shannon's source coding theorem~\cite{shannon48}, this is a contradiction.

For reasons that become apparent later, our encoding and decoding procedures will share a random source and also both need access to $x_{n^2},\dots,x_{\beta^\ell}$. The randomness we need is a list $\Gamma_1,\dots,\Gamma_m$ of $m=|\F|^{nk/512}$ independently chosen sets, where each $\Gamma_i$ is a uniform random set of $k=(\beta^\ell -1 )/n$ vectors from $\F^n$. Observe that $x_{\beta^\ell-1} \cdots x_1$ are independent of $x_{n^2} \cdots x_{\beta^\ell}$ and $\Gamma_1 \cdots \Gamma_m$, thus $H(x_{\beta^{\ell}-1} \cdots x_1 \mid x_{n^2} \cdots x_{\beta^\ell} \Gamma_1 \cdots \Gamma_m) = \beta^\ell \lg |\F|$ and we still reach a contradiction if we are able to encode $x_{\beta^\ell-1} \cdots x_1$ in less than $\beta^\ell \lg |\F|$ bits in expectation when the encoder and decoder share $x_{n^2}\cdots x_{\beta^\ell}$ and $\Gamma_1 \cdots \Gamma_m$.

The encoding argument will show that, assuming~\eqref{eqn:assump}, we can often find a small set of queries, all probing the same small set of cells, and that collectively reveal a lot of information about the updates of epochs $\ell,\dots,1$. To this end, we need to formalize exactly how these queries reveal a lot of information. We thus need a few definitions:
\begin{definition}
Let $S \subseteq \{1,\dots,n\}$ be a set of indices and let $v \in \F^n$ be a vector. Then $v^{|S}$ is the vector with $|S|$ entries, one for each index $i \in S$. The coordinate in $v^{|S}$ corresponding to an index $i \in S$ has the value $v(i)$.
\end{definition}
\begin{definition}
Let $S_1,\dots,S_n \subseteq \{1,\dots,n\}$ be $n$ sets of indices and let $v_1,\dots,v_k \in \F^n$ be $k$ vectors. Then the \emph{rank sum} of $v_1,\dots,v_k$ with respect to $S_1,\dots,S_n$, denoted $\RankSum(S_1,\dots,S_n,v_1,\dots,v_k)$ is defined as
$$\RankSum(S_1,\dots,S_n,v_1,\dots,v_k) := \sum_{i=1}^n \dim(\vspan(v_1^{|S_i},\dots,v_k^{|S_i})).$$
\end{definition}
\begin{definition}
For $i=1,\dots,n$ let $R^{\leq \ell}_i$ be the set of column indices updated in the $i$'th row during epochs $\ell,\dots,1$, i.e. $R^{\leq \ell}_i = \{j_k : k \leq \beta^{\ell}-1 \wedge i_k=i\}$. Let $v_1,\dots,v_k \in \F^n$ be a set of $k$ vectors. Then the \emph{rank sum} of $v_1,\dots,v_k$ wrt. epoch $\ell$, denoted $\RankSum^{\leq \ell}(v_1,\dots,v_k)$, is defined as:
$$
\RankSum^{\leq \ell}(v_1,\dots,v_k) := \RankSum(R^{\leq \ell}_1,\dots,R^{\leq \ell}_n,v_1,\dots,v_k).
$$
\end{definition}
Note that $R^{\leq \ell}_i$ is \emph{not} random since we always update the same fixed sequence of indices $(i_k,j_k)$, it is only the value $x_k$ that varies in our hard distribution. With these definitions, it should be intuitive that a set of query vectors $v_1,\dots,v_k$ and their corresponding answers $Mv_1,\dots,Mv_k$ will reveal a lot of information about epochs $\ell,\dots,1$ if $\RankSum^{\leq \ell}(v_1,\dots,v_k)$ is high.  To exploit this in an encoding argument, we show that assumption~\eqref{eqn:assump} implies that we can find a small subset of cells in $C_\ell(\Pi)$ that answers a set of queries with high rank sum. The precise details are as follows
\begin{lemma}
\label{lem:cellset}
Let $\ell \geq (4/6)\lg n^2$ and assume~\eqref{eqn:assump}. Then with probability at least $1/4$ over the choice of $\Pi$, there exists a subset of cells $C^*_\ell(\Pi) \subseteq C_\ell(\Pi)$ satisfying:
\begin{enumerate}
\item $|C^*_\ell(\Pi)| = \beta^\ell \lg|\F|/(1024w)$.
\item There exists at least $|\F|^{(1-o(1))nk}$ distinct sets of $k=(\beta^\ell-1)/n$ query vectors $v_1,\dots,v_k$ for which:
\begin{enumerate}
\item $D^*$ does not err when answering $v_i$ after the updates $\Pi$ for $i=1,\dots,k$.
\item $P(\Pi,v_i) \cap (C_\ell(\Pi) \setminus C^*_\ell(\Pi))=\emptyset$ for $i=1,\dots,k$.
\item $\RankSum^{\leq \ell}(v_1,\dots,v_k) \geq nk/32$.
\end{enumerate}
\end{enumerate}
\end{lemma}
We briefly discuss the main intuition on why Lemma~\ref{lem:cellset} eventually leads to a contradiction to assumption~\eqref{eqn:assump}: Assuming~\eqref{eqn:assump}, the lemma says that for most outcomes of $\Pi$, one can find a relatively small subset $C^*_\ell(\Pi)$ of cells in $C_\ell(\Pi)$, where there is a large number of queries that read nothing else from epoch $\ell$ than the cells in $C^*_\ell(\Pi)$ (property 2.b). These queries must intuitively ``collect'' the information they need about epoch $\ell$ from this small set of cells. But property 2.c says that they need a lot of information, in fact even more than the bits in $C^*_\ell(\Pi)$ can possibly describe. This is the high level message of the lemma and eventually gives the contradiction.

To not remove focus from bounding the probes to epoch $\ell$, we defer the proof of Lemma~\ref{lem:cellset} to Section~\ref{sec:cellset} and instead show how we use it in the encoding argument. So let $\ell \geq (4/6)\lg n^2$ and assume~\eqref{eqn:assump}.
Under this assumption, we show how to encode and decode $x_{\beta^{\ell}-1} \cdots x_1$ in less than $H(x_{\beta^{\ell}-1} \cdots x_1 \mid x_{n^2} \cdots x_{\beta^\ell} \Gamma_1\cdots \Gamma_m) = \beta^\ell \lg |\F|$ bits in expectation. Note that we condition on $\Gamma_1 \cdots \Gamma_m$ and $x_{n^2} \cdots x_{\beta^\ell}$, which is shared information between the encoder and decoder.
\paragraph{Encoding Procedure.}
Given $\Pi = (i_{n^2},j_{n^2},x_{n^2}),\dots,(i_1,j_1,x_1)$ to encode, first observe that the indices $i_k$ and $j_k$ are fixed, thus we only need to encode $x_{\beta^{\ell}-1} \cdots x_1$. We proceed as follows:
\begin{enumerate}
\item We start by running the updates $\Pi$ on $\D^*$. We then check if a cell set $C^*_\ell(\Pi) \subseteq C_\ell(\Pi)$ satisfying the properties in Lemma~\ref{lem:cellset} exists. If not, our encoding consists of a $0$-bit, followed by a naive encoding of $x_{\beta^\ell-1}\dots x_1$, costing $1+\beta^\ell \lg |\F|$ bits. In this case, we terminate the encoding procedure. Note that under assumption~\eqref{eqn:assump}, this happens with probability at most $3/4$.
\item If a cell set $C_\ell^*(\Pi) \subseteq C_\ell(\Pi)$ satisfying the properties in Lemma~\ref{lem:cellset} does exists, we let $\Fam$ denote the family of all sets of $k=(\beta^\ell-1)/n$ query vectors satisfying 2.a-c in Lemma~\ref{lem:cellset}. We have $|\Fam| \geq |\F|^{(1-o(1))nk}$. We then check if one of the sets in $\Fam$ equals one of the random sets $\Gamma_1,\dots,\Gamma_m$. If not, we also write a $0$-bit followed by a naive encoding of $x_{\beta^\ell-1}\dots x_1$ and terminate the encoding procedure. This costs $1+\beta^\ell \lg |\F|$ bits. Recalling that $\Gamma_1,\dots,\Gamma_m$ are chosen independently of $\Pi$, we conclude that the probability of using a naive encoding of $x_{\beta^\ell-1}\cdots x_1$ in either step 1. or 2. is bounded by $3/4+(1-|\Fam|/\binom{|\F|^n}{k})^m \leq  3/4 + \exp(-m|\Fam|/|\F|^{nk}) \leq 3/4 + \exp(-|\F|^{\Omega(nk)}) \leq 4/5$.
\item If we did not terminate and write a naive encoding in either step 1. or 2. above, we have found a $C^*_\ell(\Pi)$ satisfying the properties in Lemma~\ref{lem:cellset} as well as an index $i^*$ amongst $\{1,\dots,m\}$ such that the vectors in $\Gamma_{i^*}$ satisfy properties 2.a-c in Lemma~\ref{lem:cellset}. We use $\gamma_1,\dots,\gamma_k$ to denote these vectors. We now write a $1$-bit, followed by an encoding of $i^*$ and the addresses and contents of cells in $C^*_\ell(\Pi)$. This costs $1 + \lg m + |C^*_\ell(\Pi)|2w \leq 1 + nk\lg|\F|/512 + nk\lg|\F|/512 \leq 1+\beta^\ell\lg|\F|/256$ bits.
\item We then write down the addresses and contents of all cells in $C_{\ell-1}(\Pi),\dots,C_1(\Pi)$. Since the worst case update time is $t_u$ and we chose $\beta = 1024t_u w/\lg|\F|$, this costs no more than $\sum_{j=1}^{\ell-1} |C_j(\Pi)|2w \leq 4\beta^{\ell-1}t_u w \leq \beta^\ell \lg |\F|/256$ bits.
\item In the last step, we iterate through the rows of $M$ from $1$ to $n$. For row $i$, we create an initially empty set $X_i$ of vectors in $\F^{|R_i^{\leq \ell}|}$. We now iterate through all vectors in $\F^{|R_i^{\leq \ell}|}$ in some arbitrary but fixed order. For each such vector $v$, we check whether $v$ is in $\vspan(\gamma_1^{|R_i^{\leq \ell}},\dots,\gamma_k^{|R_i^{\leq \ell}},X_i)$. If not, we add $v$ to $X_i$. We then continue to the next vector in $\F^{|R_i^{\leq \ell}|}$. Once this terminates, we have $\dim(\vspan(\gamma_1^{|R_i^{\leq \ell}},\dots,\gamma_k^{|R_i^{\leq \ell}},X_i) = |R_i^{\leq \ell}|$ and $|X_i| = |R_i^{\leq \ell}| - \dim(\vspan(\gamma_1^{|R_i^{\leq \ell}},\dots,\gamma_k^{|R_i^{\leq \ell}}))$. Letting $m_i$ denote the $i$'th row vector in $M$ after having processed all the updates $\Pi$, we finally compute $\langle m_i^{|R_i^{\leq \ell}}, v \rangle$ for each $v \in X_i$. We write down these inner products, costing $|X_i|\lg |\F|$ bits. Summing over all rows $i$, this step costs $\sum_i |X_i|\lg|\F| = \sum_i (|R_i^{\leq \ell}| - \dim(\vspan(\gamma_1^{|R_i^{\leq \ell}},\dots,\gamma_k^{|R_i^{\leq \ell}})))\lg|\F| = (\beta^\ell - \RankSum^{\leq \ell}(\gamma_1,\dots,\gamma_k))\lg |\F| \leq (\beta^\ell - nk/32)\lg|\F| = (\beta^\ell-(\beta^\ell-1)/32)\lg|\F|$ bits.
\end{enumerate}
Examining the above encoding procedure, we see that the expected length of the encoding is upper bounded by
\begin{eqnarray*}
1 + (4/5)(\beta^\ell \lg|\F|) + (1/5)(\beta^\ell \lg|\F| + \beta^\ell \lg|\F|/128 - (\beta^\ell-1)\lg|\F|/32) &\leq&\\
1 + \beta^\ell \lg|\F| - (1/5)(3\beta^\ell \lg|\F|/128 - \lg|\F|/32) &=&\\
H(x_{\beta^\ell - 1} \cdots x_1 \mid x_{n^2} \cdots x_{\beta^\ell} \Gamma_1 \cdots \Gamma_m) - \Omega(\beta^\ell \lg |\F|) &<&\\
H(x_{\beta^\ell - 1} \cdots x_1 \mid x_{n^2} \cdots x_{\beta^\ell} \Gamma_1 \cdots \Gamma_m).
\end{eqnarray*}
Thus to reach the contradiction to assumption~\eqref{eqn:assump}, we only need to show that we can recover $x_{\beta^\ell-1} \cdots x_1$ from the above encoding and $x_{n^2}\cdots x_{\beta^\ell} \Gamma_1 \cdots \Gamma_m$. We do this as follows:
\paragraph{Decoding Procedure.}
\begin{enumerate}
\item First we check the first bit of the encoding. If this is a $0$-bit, we recover $x_{\beta^\ell-1} \cdots x_1$ directly from the remaining part of the encoding and terminate.
\item If the first bit is a $1$-bit, we first recover $i^*, C^*_\ell(\Pi)$ and $C_{\ell-1}(\Pi),\dots,C_1(\Pi)$. From $i^*$ and $\Gamma_1,\dots,\Gamma_m$ we also recover $\gamma_1,\dots,\gamma_k$. Since the update indices $i_{n^2},\dots,i_1$ and $j_{n^2},\dots,j_1$ are fixed, we can also compute $\gamma_1^{|R^{\leq \ell}_i},\dots,\gamma_k^{|R^{\leq \ell}_i}$ for all rows $i$.
\item For each row $i$ in turn, we create an initially empty set $X_i$ of vectors in $\F^{|R_i^{\leq \ell}|}$. We then iterate through all vectors $v \in \F^{|R_i^{\leq \ell}|}$ in the same fixed order as in step 5. of the encoding procedure. For each such $v$, we check if $v$ is in $\vspan(\gamma_1^{|R_i^{\leq \ell}},\dots,\gamma_k^{|R_i^{\leq \ell}},X_i)$. If not, we add $v$ to $X_i$. We then continue to the next $v$. When this terminates we have reconstructed the sets $X_i$ for all rows $i$.
\item We now process the updates $(i_{n^2},j_{n^2},x_{n^2}),\dots,(i_{\beta^\ell},j_{\beta^\ell}, x_{\beta^\ell})$ on $D^*$, that is, we process all updates until just before epoch $\ell$. We have thus computed the contents of every memory cell at the time just before epoch $\ell$.
\item We now run the query algorithm of $D^*$ for $\gamma_1,\dots,\gamma_k$. When answering the query $\gamma_i$, the query algorithm repeatedly asks for a memory cell. When asking for a memory cell, we first check if the cell is amongst $C_{\ell-1}(\Pi),\dots,C_1(\Pi)$. If so, we have its contents and can continue to the next probe. Otherwise we check if the cell is amongst $C^*_{\ell}(\Pi)$. If so, we again have its contents and can continue to the next probe. If not, we know by property 2.b of Lemma~\ref{lem:cellset} that the cell is not in $C_{\ell}(\Pi)$, i.e. it was not updated during epochs $\ell,\dots,1$. Thus its contents after processing all of $\Pi$ is the same as after processing only updates preceding epoch $\ell$ and we thus have its contents from the previous step of the decoding procedure (step 4.). Since we are able to run the entire query algorithm, we get from property 2.a in Lemma~\ref{lem:cellset} that we recover the vector $M\gamma_i$ for each $i=1,\dots,k$.
\item Finally, for each row $i=1,\dots,n$ in $M$, we will recover all the values $x_h$ for which $h \leq \beta^{\ell}-1$ and $i_h = i$. This is precisely the values corresponding to the updates of the columns given by $R_i^{\leq \ell}$. We do this as follows: First, from the query answer $M \gamma_h$ for $h=1,\dots,k$ we can directly read off $\langle m_i , \gamma_h \rangle$ where $m_i$ is the $i$'th row of $M$ after processing $\Pi$. Since the decoder is given access to $x_{n^2},\dots,x_{\beta^\ell}$, this allows us to compute $\langle m_i^{|R^{\leq \ell}_i} , \gamma_h^{|R^{\leq \ell}_i} \rangle$. Finally from step 5. of the encoding procedure we also have $\langle m_i^{|R^{\leq \ell}_i} , v \rangle$ for every $v \in X_i$. Since $\dim(\vspan(\gamma_1^{|R_i^{\leq \ell}},\dots,\gamma_k^{|R_i^{\leq \ell}},X_i) = |R_i^{\leq \ell}|$, this uniquely determines $m_i^{|R^{\leq \ell}_i}$ and thus the values $x_h$ for which $h \leq \beta^\ell - 1$ and $i_j = i$. Doing this for all rows finally recovers $x_{\beta^{\ell}-1},\dots,x_1$.
\end{enumerate}

To summarize, we showed how to encode and decode $x_{\beta^{\ell}-1},\dots,x_1$ in less than $H(x_{\beta^{\ell}-1}\cdots x_1)$ bits under assumption~\eqref{eqn:assump}. This is a contradiction, completing the proof of Lemma~\ref{lem:epoch}.

\subsection{Finding a Cell Set (Proof of Lemma~\ref{lem:cellset})}
\label{sec:cellset}
In this section, we prove Lemma~\ref{lem:cellset}. For this, first define $W(\Pi)$ as the set of query vectors $w$ for which $D^*$ does not err when answering $w$ after processing $\Pi$, and at the same time, 
$$|C_\ell(\Pi) \cap P(\Pi,w)|  \leq 16 \E_{\Pi,v}[|C_\ell(\Pi) \cap P(\Pi,v)|].$$ 
Define $\Err(\Pi,v)$ as the indicator random variable taking the value $1$ if $D^*$ errs when answering $v$ after updates $\Pi$ and $0$ otherwise. We have $\E_{\Pi,v}[\Err(\Pi,v)] \leq 1/3$. By Markov's inequality and a union bound, we conclude that with probability at least $1/4$ over the choice of $\Pi$, we have both $\E_v[|C_\ell(\Pi) \cap P(\Pi,v)|] \leq 4 \E_{\Pi,v}[|C_\ell(\Pi) \cap P(\Pi,v)|]$ and $\E_v[\Err(\Pi,v)] \leq 2/3$. We say that the update sequence $\Pi$ is \emph{good} when this happens. We can again use Markov's inequality and a union bound to conclude that $|W(\Pi)| \geq |\F|^n/12$ when $\Pi$ is good. Now consider a vector $w \in W(\Pi)$ and let $\Delta=\beta^\ell \lg |\F|/(1024w)$. Observe that there are $\binom{|C_\ell(\Pi)| - | C_\ell(\Pi) \cap P(\Pi,w)|}{\Delta-|C_\ell(\Pi) \cap P(\Pi,w)|}$ subsets $C' \subseteq C_\ell(\Pi)$ of $\Delta$ cells, satisfying $P(\Pi,w) \cap (C_\ell(\Pi) \setminus C') = \emptyset$. By averaging, this means that there is a set $C^*_\ell(\Pi) \subseteq C_\ell(\Pi)$ of $\Delta$ cells, with at least $|W(\Pi)|\binom{|C_\ell(\Pi)| - | C_\ell(\Pi) \cap P(\Pi,w)|}{\Delta-|C_\ell(\Pi) \cap P(\Pi,w)|}/\binom{|C_\ell(\Pi)|}{\Delta}$ distinct vectors $w \in W(\Pi)$ satisfying $P(\Pi,w) \cap (C_\ell(\Pi) \setminus C^*_\ell(\Pi)) = \emptyset$. This is lower bounded by
\begin{eqnarray*}
\frac{|W(\Pi)| \binom{|C_\ell(\Pi)| - | C_\ell(\Pi) \cap P(\Pi,w)|}{\Delta-|C_\ell(\Pi) \cap P(\Pi,w)|}}{\binom{|C_\ell(\Pi)|}{\Delta}} &=& \frac{|W(\Pi)|(|C_\ell(\Pi)| - | C_\ell(\Pi) \cap P(\Pi,w)|)!\Delta!(|C_\ell(\Pi)|-\Delta)!}{|C_\ell(\Pi)|!(\Delta-|C_\ell(\Pi) \cap P(\Pi,w)|)!(|C_\ell(\Pi)|-\Delta)!}\\
&\geq&  \frac{|W(\Pi)|(\Delta-|C_\ell(\Pi) \cap P(\Pi,w)|)^{|C_\ell(\Pi) \cap P(\Pi,w)|}}{|C_\ell(\Pi)|^{|C_\ell(\Pi) \cap P(\Pi,w)|}}.
\end{eqnarray*}
When $\Pi$ is good, assumption~\eqref{eqn:assump} implies $|C_\ell(\Pi) \cap P(\Pi,w)| = o(\beta^{\ell} \lg |\F|/w)$ and we chose $\Delta=\beta^\ell \lg|\F|/(1024 w)$. Thus for good $\Pi$, the above is at least
\begin{eqnarray*}
&\geq&  |W(\Pi)|\left( \frac{\Delta}{2|C_\ell(\Pi)|} \right)^{|C_\ell(\Pi) \cap P(\Pi,w)|}\\
&\geq& |W(\Pi)|\left( \frac{\Delta}{2\beta^\ell t_u} \right)^{|C_\ell(\Pi) \cap P(\Pi,w)|}\\
&=& |W(\Pi)| \left( \frac{\lg|\F|}{2048t_u w} \right)^{|C_\ell(\Pi) \cap P(\Pi,w)|}.
\end{eqnarray*}
For good $\Pi$, assumption~\eqref{eqn:assump} also implies $|C_\ell(\Pi) \cap P(\Pi,w)| = o(n \lg|\F|/(\lg((t_u w)/\lg|\F|)))$. Inserting this above we get
\begin{eqnarray*}
&\geq& |W(\Pi)| |\F|^{-o(n)}\\
&=& |\F|^{(1-o(1))n}.
\end{eqnarray*}
Thus for good $\Pi$, we have at least $|\F|^{(1-o(1))n}$ vectors $w$ such that $D^*$ does not err when answering $w$ after processing $\Pi$ and also $P(\Pi,w) \cap (C_\ell(\Pi) \setminus C^*_\ell(\Pi)) = \emptyset$. Let the set of these vectors be denoted $U(\Pi)$. Now let $k=(\beta^{\ell}-1)/n$ and consider all $k$-sized subsets of $U(\Pi)$. There are $\binom{|U(\Pi)|}{k} = |\F|^{(1-o(1))nk}$ distinct such sets. We want to show that most of these sets have high rank sum. For this, we have the following lemma:
\begin{lemma}
\label{lem:highranksum}
If $\ell \geq (4/6)\lg_\beta n^2$ and $k=(\beta^\ell -1)/n$, then there exists no more than $|\F|^{(27/32) nk}$ distinct sets of $k$ vectors, $v_1,\dots,v_k$ in $\F^n$, such that $\RankSum^{\leq \ell}(v_1,\dots,v_k) < nk/32$.
\end{lemma}
Observe that combined with $|U(\Pi)| = |\F|^{(1-o(1))nk}$ for good $\Pi$ and $\Pr[\Pi \textrm{ is good}] \geq 1/4$, Lemma~\ref{lem:highranksum} immediately implies Lemma~\ref{lem:cellset}. We thus finish our proof of Lemma~\ref{lem:cellset} by proving Lemma~\ref{lem:highranksum}.

\paragraph{Vectors and Rank Sum (Proof of Lemma~\ref{lem:highranksum}).}
In the following, we prove Lemma~\ref{lem:highranksum}. Let $\ell \geq (4/6)\lg_\beta n^2$ and let $V$ be the family of all sets of $k=(\beta^\ell-1)/n$ distinct vectors in $\F^n$ for which $\RankSum^{\leq \ell}(v_1,\dots,v_k) < nk/32$. Our goal is to show that $V$ has to be small. The intuition for why this is true is as follows: If a set of vectors $v_1,\dots,v_k$ has small rank sum wrt. epoch $\ell$, then for most rows $i$, we have that $\dim(\vspan(v_1^{|R_i^{\leq \ell}},\dots,v_k^{|R_i^{\leq \ell}}))$ is small. This means that, when restricted to the columns in $R_i^{\leq \ell}$, the vectors $v_1,\dots,v_k$ must be contained in a low dimensional space. Since there are not too many vectors in a low dimensional space, this gives a bound on the size of $V$ when restricted to the coordinates in $R_i^{\leq \ell}$. From there, our choice of well-separated update indices also comes into play. This property of the update indices basically ensures that different rows put constraints on different coordinates of $v_1,\dots,v_k$, effectively ensuring that if $v_1,\dots,v_k$ is a set in $V$, then they must lie in a low dimensional subspace no matter which subset of columns we consider. This finally gives a bound on $|V|$. We formalize this intuition using an encoding argument.

\paragraph{Encoding Argument.}
Let $v_1,\dots,v_k$ be a set of $k=(\beta^\ell-1)/n$ vectors from the family $V$. We present an efficient encoding and decoding procedure for $v_1,\dots,v_k$. This gives a bound on $|V|$. The encoding procedure is as follows:
\begin{enumerate}
\item Given $v_1,\dots,v_k$, we let $I$ be the set of all row indices $i$ for which $\dim(\vspan(v_1^{|R_i^{\leq \ell}},\dots,v_k^{|R_i^{\leq \ell}})) \leq k/16$. Since we assumed $\RankSum^{\leq \ell}(v_1,\dots,v_k) < nk/32$ and by definition $\RankSum^{\leq \ell}(v_1,\dots,v_k) = \RankSum(R_1^{\leq \ell},\dots,R_n^{\leq \ell})$, it follows from Markov's inequality that $|I| \geq n/2$.
\item Since our update indices $(i_h,j_h)$ were chosen as well-spread, and since $\ell \geq   (4/6)\lg_\beta n^2$, it follows from Definition~\ref{def:spread} that we can find a subset $I^* \subseteq I$ of indices satisfying $|I^*| \leq 8n^2/(\beta^\ell-1) = 8n/k$ and $\left|\bigcup_{h \leq \beta^\ell-1 : i_h \in I^*} \{j_h\} \right| \geq n/4$. The first part of our encoding is such a set of indices $I^*$. This costs no more than $8n \lg n/k$ bits.
\item For each $i \in I^*$, in increasing order, let $\tilde{R}_i^{\leq \ell} = R_i^{\leq \ell} \setminus \left\{\bigcup_{i' \in I^* : i' < i} R_{i'}^{\leq \ell} \right\}$. If $\tilde{R}_i^{\leq \ell}$ is empty, we continue to the next index in $I^*$. Otherwise, find some basis $w_1,\dots,w_{d}$ for $\vspan(v_1^{|\tilde{R}_i^{\leq \ell}},\dots,v_k^{|\tilde{R}_i^{\leq \ell}})$. Observe that $d \leq k/16$ since $I^* \subseteq I$. We write down $d$ and $w_1,\dots,w_d$. This costs no more than $\lg n + d|\tilde{R}_i^{\leq \ell}|\lg|\F|$ bits. After having specified $w_1,\dots,w_d$, we also write down $\langle w_j, v_h^{|\tilde{R}_i^{\leq \ell}} \rangle$ for each pair $j \in \{1,\dots,d\}$ and $h \in \{1,\dots,k\}$, costing $dk\lg|\F|$ bits. Summing over all $i \in I^*$, the total cost of this step is thus at most $|I^*|\lg n + (k/16)\lg|\F| \sum_{i \in I^*}|\tilde{R}_i^{\leq \ell}| + |I^*|(k/16)k\lg|\F|$ bits. This is at most $8n\lg n/k +  (k/16)\lg|\F| \sum_{i \in I^*}|\tilde{R}_i^{\leq \ell}| + nk\lg|\F|/16$.
\item Finally we let $X$ be the set of column indices not contained in any $R_i^{\leq \ell}$ with $i \in I^*$. For each vector $v_1,\dots,v_k$ in turn, we write down each coordinate corresponding to a column in $X$. This costs $k|X|\lg|\F|$ bits.
\end{enumerate}
Next we show that we can recover $v_1,\dots,v_k$ from the encoding produced by the above procedure. This is done as follows:
\begin{enumerate}
\item From the bits written in step 2. of the encoding procedure, we recover $I^*$.
\item For each $i \in I^*$, in increasing order, compute $\tilde{R}_i^{\leq \ell} = R_i^{\leq \ell} \setminus \left\{\bigcup_{i' \in I^* : i' < i} R_{i'}^{\leq \ell} \right\}$ (these sets depend only on $i$ and the fixed update indices). If $\tilde{R}_i^{\leq \ell}$ is empty, we continue with the next index in $I^*$. If not, we read the value $d$ and the basis $w_1,\dots,w_d$ written for this index $i \in I^*$. We then read $\langle w_j, v_h^{|\tilde{R}_i^{\leq \ell}} \rangle$ for each pair $j \in \{1,\dots,d\}$ and $h \in \{1,\dots,k\}$. Since each $v_h^{|\tilde{R}_i^{\leq \ell}}$ is in $\vspan(v_1^{|\tilde{R}_i^{\leq \ell}},\dots,v_k^{|\tilde{R}_i^{\leq \ell}})$, these inner products allows us to recover each coordinate  $v_h(c)$ for any column index $c$ in $\tilde{R}_i^{\leq \ell}$ and any $h=1,\dots,k$.
\item What remains is to recover all coordinates corresponding to column indices $c$ where $c \notin \tilde{R}_i^{\leq \ell}$ for any $h=1,\dots,k$. But $\bigcup_{i \in I^*}\tilde{R}_i^{\leq \ell} = \bigcup_{i \in I^*} R_i^{\leq \ell}$ and thus we recover the remaining coordinates from the bits written in step 4. of the encoding procedure.
\end{enumerate}
We have thus shown that we can encode and decode every set of $k$ vectors $v_1,\dots,v_k$ into a string of at most:
$$
16n\lg n/k + (k/16)\lg|\F| \sum_{i \in I^*}|\tilde{R}_i^{\leq \ell}| + nk\lg|\F|/16 + k|X|\lg|\F|
$$
bits. But $\sum_{i \in I^*}|\tilde{R}_i^{\leq \ell}| = n-|X|$, and we rewrite $|X| = n-\sum_{i \in I^*}|\tilde{R}_i^{\leq \ell}|$. The above is thus equal to
$$
nk\lg|\F| + 16n\lg n/k + nk\lg|\F|/16 -(15/16)k \lg|\F| \sum_{i \in I^*}|\tilde{R}_i^{\leq \ell}|.
$$
We also have $\sum_{i \in I^*}|\tilde{R}_i^{\leq \ell}| = \left|\bigcup_{h \leq \beta^{\ell}-1 : i_h \in I^*} \{j_h\} \right| \geq n/4$. This means that our encoding uses no more than
\begin{eqnarray*}
nk\lg|\F| + 16n\lg n/k + nk\lg|\F|/16 -(15/64)nk \lg|\F| &=&\\
(53/64)nk\lg|\F|+ 16n\lg n/k &<&\\
(54/64)nk\lg|\F|
\end{eqnarray*}
bits. We thus conclude $|V| \leq 2^{(27/32) nk \lg|\F|}$.
\subsection{Well-Spread Sequence (Proof of Lemma~\ref{lem:existspread})}
\label{sec:spread}
Let $r$ be some value in the range $n^{4/3} \leq r \leq n^2/8$ and let $S$ be a set $n/2$ row indices. Partition $S$ into $|S|/(n^2/r) = r/(2n)$ consecutive groups $G_1,\dots,G_{r/(2n)}$ of $n^2/r$ rows each. Now let  $(i_{n^2},j_{n^2}),\dots,(i_1,j_1)$ be a uniform random permutation of the $n^2$ pairs of indices in $\{1,\dots,n\} \times \{1,\dots, n\}$. We bound the probability that $\left|\bigcup_{k \leq r : i_k \in G_h} \{j_k\}\right| < n/4$ for all $h=1,\dots,r/(2n)$. This probability is upper bounded by
\begin{equation*}
\frac{\binom{n}{n/2}^{r/(2n)} \binom{5n^2/8}{r} }{\binom{n^2}{r}}.
\end{equation*}
To see why, observe that if $\left|\bigcup_{k \leq r : i_k \in G_h} \{j_k\}\right| < n/4$ for all $h=1,\dots,r/(2n)$, then for all $h=1,\dots,r/(2n)$, there must exist a set of $3n/4$ column indices $C_h$ such that $\left(\bigcup_{k \leq r} \{(i_k,j_k)\}\right) \bigcap \left(G_h \times C_h\right) = \emptyset$. Thus we must have:
\begin{equation}
\label{eq:disj}
\left(\bigcup_{k \leq r} \{(i_k,j_k)\}\right) \bigcap \left(\bigcup_{h=1}^{r/(2n)} \left(G_h \times C_h\right)\right) = \emptyset
\end{equation}
At the same time, we have
$$
\left|\bigcup_{h=1}^{r/(2n)} \left(G_h \times C_h\right)\right| = (r/(2n))(n^2/r)(3n/4) = 3n^2/8.
$$
These inequalities explain our upper bound. The $\binom{n}{n/4}^{r/(2n)}$ term counts the number of possible families of sets $C_1,\dots,C_h$ that could satisfy \eqref{eq:disj}. For a particular choice of $C_1,\dots,C_h$, there are only $\binom{5n^2/8}{r}$ choices of $\bigcup_{k \leq r} \{(i_k,j_k)\}$ that gives the required disjointness. The denominator just counts the total number of choices of $\bigcup_{k \leq r} \{(i_k,j_k)\}$. We continue our calculations:
\begin{eqnarray*}
\frac{\binom{n}{n/4}^{r/(2n)} \binom{5n^2/8}{r} }{\binom{n^2}{r}} &\leq& (4e)^{r/8}\frac{(5n^2/8)!r!(n^2-r)!}{(n^2)!r!(5n^2/8-r)!} \\
&\leq& (4e)^{r/8} \frac{(5n^2/8)^r}{(n^2-r+1)^r}.
\end{eqnarray*}
Since $r \leq n^2/8$, this is at most
\begin{eqnarray*}
(4e)^{r/8} \frac{(5n^2/8)^r}{(7n^2/8)^r} &<& 1.35^{r} \left(\frac{5}{7}\right)^r \\
&<& 0.97^r\\
&=& 2^{-\Omega(n^{4/3})}.
\end{eqnarray*}
Thus for a particular choice of $n^{4/3} \leq r \leq n^2/8$ and set $S$ of $n/2$ rows, the probability that there does not exist a subset $S^* \subseteq S$ with $|S^*| \leq n^2/r$ and $\left|\bigcup_{k \leq r : i_k \in S^*} \{j_k\}\right|\geq n/4$ is at most $2^{-\Omega(n^{4/3})}$. Since there are less than $n^2 \binom{n}{n/2} < 2^{2n}$ possible choices for $r$ and $S$, we can union bound over all of them and conclude that there exists a sequence of update indices $(i_{n^2},j_{n^2}),\dots,(i_1,j_1)$ such that for any $n^{4/3} \leq r \leq n^2/8$ and any set $S$ of $n/2$ rows, there exists a subset $S^* \subseteq S$ with size $n^2/r$ satisfying $\left|\bigcup_{k \leq r : i_k \in S^*} \{j_k\}\right|\geq n/4$. For the case $n^2/8 < r \leq n^2$ and any set $S$ of $n/2$ rows, the same sequence of updates must necessarily have a subset $S^*$ of size at most $8n^2/r$ for which $\left|\bigcup_{k \leq r : i_k \in S^*} \{j_k\}\right|\geq n/4$. Thus $(i_{n^2},j_{n^2}),\dots,(i_1,j_1)$ is well-spread.
 
\bibliographystyle{plain}
\bibliography{longnames.bib,bib-latest.bib,bristol.bib}

\begin{thebibliography}{10}

\bibitem{AW:2014}
Amir Abboud and Virginia~Vassilevska Williams.
\newblock Popular conjectures imply strong lower bounds for dynamic problems.
\newblock In {\em FOCS '14: Proc. 55\textsuperscript{th} Annual Symp.
  Foundations of Computer Science}, 2014.

\bibitem{AWW:2014}
Amir Abboud, Virginia~Vassilevska Williams, and Oren Weimann.
\newblock Consequences of faster alignment of sequences.
\newblock In {\em ICALP '14: Proc. 41\textsuperscript{st} International
  Colloquium on Automata, Languages and Programming}, pages 39--51, 2014.

\bibitem{BI:SETH:2015}
Arturs Backurs and Piotr Indyk.
\newblock Edit distance cannot be computed in strongly subquadratic time
  (unless {SETH} is false).
\newblock In {\em STOC '15: Proc. 47\textsuperscript{th} Annual ACM Symp.
  Theory of Computing}, 2015.

\bibitem{Bringmann:2014}
Karl Bringmann.
\newblock Why walking the dog takes time: Frechet distance has no strongly
  subquadratic algorithms unless {SETH} fails.
\newblock In {\em FOCS '14: Proc. 55\textsuperscript{th} Annual Symp.
  Foundations of Computer Science}, pages 661--670, 2014.

\bibitem{BK:TimeWarping:2015}
Karl Bringmann and Marvin K{\"{u}}nnemann.
\newblock Quadratic conditional lower bounds for string problems and dynamic
  time warping.
\newblock {\em CoRR}, abs/1502.01063, 2015.

\bibitem{CJS-soda:2015}
R.~Clifford, M.~Jalsenius, and B.~Sach.
\newblock Cell-probe bounds for online edit distance and other pattern matching
  problems.
\newblock In {\em SODA '15: Proc. 26\textsuperscript{th} ACM-SIAM Symp. on
  Discrete Algorithms}, 2015.

\bibitem{CJ:2011}
Rapha\"{e}l Clifford and Markus Jalsenius.
\newblock Lower bounds for online integer multiplication and convolution in the
  cell-probe model.
\newblock In {\em ICALP '11: Proc. 38\textsuperscript{th} International
  Colloquium on Automata, Languages and Programming}, pages 593--604, 2011.

\bibitem{CJS:2013}
Rapha\"{e}l Clifford, Markus Jalsenius, and Benjamin Sach.
\newblock Tight cell-probe bounds for online hamming distance computation.
\newblock In {\em SODA '13: Proc. 24\textsuperscript{th} ACM-SIAM Symp. on
  Discrete Algorithms}, pages 664--674, 2013.

\bibitem{Dietzfelbinger}
Martin Dietzfelbinger, Torben Hagerup, Jyrki Katajainen, and Martti Penttonen.
\newblock A reliable randomized algorithm for the closest-pair problem.
\newblock {\em Journal of Algorithms}, 25(1):19 -- 51, 1997.

\bibitem{FHM:2001}
G.S. Frandsen, J.P. Hansen, and P.B. Miltersen.
\newblock Lower bounds for dynamic algebraic problems.
\newblock {\em Information and Computation}, 171(2):333--349, 2001.

\bibitem{Fredman:1978}
M.~Fredman.
\newblock Observations on the complexity of generating quasi-{G}ray codes.
\newblock {\em SIAM Journal on Computing}, 7(2):134--146, 1978.

\bibitem{FS1989:chronogram}
M.~Fredman and M.~Saks.
\newblock The cell probe complexity of dynamic data structures.
\newblock In {\em STOC '89: Proc. 21\textsuperscript{st} Annual ACM Symp.
  Theory of Computing}, pages 345--354, 1989.

\bibitem{GO:1995}
A.~Gajentaan and M.~H. Overmars.
\newblock On a class of ${O}(n^2)$ problems in computational geometry.
\newblock {\em Computational Geometry}, 5(3):165--185, 1995.

\bibitem{HKNS:2015}
M.~Henzinger, S.~Krinninger, D.~Nanongkai, and T.~Saranurak.
\newblock Unifying and strengthening hardness for dynamic problems via the
  online matrix-vector multiplication conjecture.
\newblock In {\em STOC '15: Proc. 47\textsuperscript{th} Annual ACM Symp.
  Theory of Computing}, 2015.

\bibitem{Larsen:2012}
Kasper~Green Larsen.
\newblock The cell probe complexity of dynamic range counting.
\newblock In {\em STOC '12: Proc. 44\textsuperscript{th} Annual ACM Symp.
  Theory of Computing}, pages 85--94, 2012.

\bibitem{Larsen:2012:focs}
Kasper~Green Larsen.
\newblock Higher cell probe lower bounds for evaluating polynomials.
\newblock In {\em FOCS '12: Proc. 53\textsuperscript{rd} Annual Symp.
  Foundations of Computer Science}, pages 293--301, 2012.

\bibitem{MNSW:1998}
P.B. Miltersen, N.~Nisan, S.~Safra, and A.~Wigderson.
\newblock On data structures and asymmetric communication complexity.
\newblock {\em Journal of Computer System Sciences}, 57(1):37--49, 1998.

\bibitem{Miltersen:1994}
Peter~Bro Miltersen.
\newblock Lower bounds for union-split-find related problems on random access
  machines.
\newblock In {\em STOC '94: Proc. 26\textsuperscript{th} Annual ACM Symp.
  Theory of Computing}, pages 625--634, 1994.

\bibitem{MP:1969}
M.~Minsky and S.~Papert.
\newblock {\em Perceptrons: An Introduction to Computational Geometry}.
\newblock MIT Press, 1969.

\bibitem{PTW:2010}
Rina Panigrahy, Kunal Talwar, and Udi Wieder.
\newblock Lower bounds on near neighbor search via metric expansion.
\newblock In {\em FOCS '10: Proc. 51\textsuperscript{st} Annual Symp.
  Foundations of Computer Science}, pages 805--814, 2010.

\bibitem{Patrascu:2008:unifyingcellprobe}
Mihai P{\v a}tra{\c s}cu.
\newblock {Unifying the landscape of cell-probe lower bounds}.
\newblock In {\em FOCS '08: Proc. 49\textsuperscript{th} Annual Symp.
  Foundations of Computer Science}, pages 434--443, 2008.

\bibitem{Patrascu:2010:polylower}
Mihai P{\v a}tra{\c s}cu.
\newblock Towards polynomial lower bounds for dynamic problems.
\newblock In {\em STOC '10: Proc. 42\textsuperscript{nd} Annual ACM Symp.
  Theory of Computing}, pages 603--610, 2010.

\bibitem{patrascu10higher}
Mihai P{\v a}tra{\c s}cu and Mikkel Thorup.
\newblock Higher lower bounds for near-neighbor and further rich problems.
\newblock {\em SIAM Journal on Computing}, 39(2):730--741, 2010.
\newblock See also FOCS'06.

\bibitem{PT:stoc:2011}
Mihai P{\v a}tra{\c s}cu and Mikkel Thorup.
\newblock Don't rush into a union: take time to find your roots.
\newblock In Lance Fortnow and Salil~P. Vadhan, editors, {\em STOC '11: Proc.
  43\textsuperscript{rd} Annual ACM Symp. Theory of Computing}, pages 559--568,
  2011.

\bibitem{PW:sat:2010}
Mihai Patrascu and Ryan Williams.
\newblock On the possibility of faster {SAT} algorithms.
\newblock In {\em SODA '10: Proc. 21\textsuperscript{st} ACM-SIAM Symp. on
  Discrete Algorithms}, pages 1065--1075, 2010.

\bibitem{PD2006:Low-Bounds}
M.~P\u{a}tra\c{s}cu and E.~D. Demaine.
\newblock Logarithmic lower bounds in the cell-probe model.
\newblock {\em SIAM Journal on Computing}, 35(4):932--963, 2006.

\bibitem{PT:pred}
Mihai P\u{a}tra\c{s}cu and Mikkel Thorup.
\newblock Time-space trade-offs for predecessor search.
\newblock In {\em STOC '06: Proc. 38\textsuperscript{th} Annual ACM Symp.
  Theory of Computing}, pages 232--240, 2006.

\bibitem{RW:sparse:2013}
Liam Roditty and Virginia~Vassilevska Williams.
\newblock Fast approximation algorithms for the diameter and radius of sparse
  graphs.
\newblock In {\em STOC '13: Proc. 45\textsuperscript{th} Annual ACM Symp.
  Theory of Computing}, pages 515--524, 2013.

\bibitem{shannon48}
Claude Shannon.
\newblock A mathematical theory of communication.
\newblock {\em Bell System Technical Journal}, 27:379--423, 623--656, July,
  October 1948.

\bibitem{Yao1981:Tables}
Andrew Chi-Chih Yao.
\newblock Should tables be sorted?
\newblock {\em Journal of the ACM}, 28(3):615--628, 1981.

\end{thebibliography}
\newpage
\appendix

\end{document}